\documentclass[12pt]{article}

\usepackage{epsfig}
\usepackage{amssymb}
\usepackage{subfigure}
\usepackage{graphicx}
\usepackage{color}
\usepackage{jheppub}

\begin{document}

\baselineskip 6mm
\renewcommand{\thefootnote}{\fnsymbol{footnote}}


\newcommand{\nc}{\newcommand}
\newcommand{\rnc}{\renewcommand}



\newcommand{\tcb}{\textcolor{blue}}
\newcommand{\tcr}{\textcolor{red}}
\newcommand{\tcg}{\textcolor{green}}


\def\be{\begin{eqnarray}}
\def\ee{\end{eqnarray}}
\def\nn{\nonumber\\}


\def\ct{\cite}
\def\la{\label}
\def\eq#1{\eqref{#1}}


\def\a{\alpha}
\def\b{\beta}
\def\g{\gamma}
\def\G{\Gamma}
\def\d{\delta}
\def\D{\Delta}
\def\e{\epsilon}
\def\et{\eta}
\def\ph{\phi}
\def\Ph{\Phi}
\def\ps{\psi}
\def\Ps{\Psi}
\def\k{\kappa}
\def\l{\lambda}
\def\L{\Lambda}
\def\m{\mu}
\def\n{\nu}
\def\th{\theta}
\def\Th{\Theta}
\def\r{\rho}
\def\s{\sigma}
\def\S{\Sigma}
\def\ta{\tau}
\def\o{\omega}
\def\O{\Omega}
\def\pr{\prime}


\def\half{\frac{1}{2}}
\def\goto{\rightarrow}

\def\na{\nabla}
\def\grad{\nabla}
\def\curl{\nabla\times}
\def\div{\nabla\cdot}
\def\pa{\partial}
\def\fr{\frac}

\def\bra{\left\langle}
\def\ket{\right\rangle}
\def\lb{\left[}
\def\lc{\left\{}
\def\ls{\left(}
\def\lp{\left.}
\def\rp{\right.}
\def\rb{\right]}
\def\rc{\right\}}
\def\rs{\right)}

\def\vac#1{\mid #1 \rangle}


\def\td#1{\tilde{#1}}
\def\check{ \maltese {\bf Check!}}


\def\Tr{{\rm Tr}\,}
\def\det{{\rm det}}
\def\text#1{{\rm #1}}


\def\bc#1{\nnindent {\bf $\bullet$ #1} \\ }
\def\ch {$<Check!>$ }
\def\ss {\vspace{1.5cm}}
\def\inf{\infty}

\begin{titlepage}

\hfill\parbox{2cm} { }

 
\vspace{2cm}

\begin{center}
{\Large \bf Holographic time-dependent entanglement entropy \\
in $p$-brane gas geometries}

\vskip 1. cm
   {Chanyong Park$^{a}$\footnote{e-mail : cyong21@gist.ac.kr}}

\vskip 0.5cm

{\it $^a$ Department of Physics and Photon Science, Gwangju Institute of Science and Technology,  Gwangju  61005, Korea}

\end{center}

\thispagestyle{empty}

\vskip3cm


\centerline{\bf ABSTRACT} \vskip 4mm

We study the time evolution of entanglement entropy in expanding universes with various matters. To describe expanding universes holographically, we take into account a braneworld moving in an asymptotic AdS space involving a uniform $p$-brane gas. In the braneworld model, an observer living in the braneworld detects the bulk motion of the braneworld as an expanding universe. We show that the entanglement entropy of expanding universes increases by the volume law in the early time and by the area law in the late time. We further consider the cosmological horizon, which is the border of the visible and invisible universe, and then investigate the time-dependent quantum entanglement between them across the cosmological horizon.

\vspace{2cm}

\end{titlepage}

\renewcommand{\thefootnote}{\axrabic{footnote}}
\setcounter{footnote}{0}



\section{Introduction}

Recently, people have paid attention to the AdS/CFT correspondence or holography for looking into nonperturbative features of a strongly interacting system \cite{Maldacena:1997re,Gubser:1998bc,Witten:1998qj,Witten:1998zw,Aharony:1999ti}. The holographic technique was further exploited to account for nontrivial quantum nature like entanglement of a ground state \cite{Ryu:2006ef,Ryu:2006bv,Nishioka:2009un,Calabrese:2009qy,Calabrese:2004eu}. Based on the AdS/CFT correspondence, in this work, we discuss the holographic dual of expanding universes with various matters and investigate the time evolution of the entanglement entropy across the cosmological (or particle) horizon.

Although the concept of the entanglement entropy is manifest in a quantum field theory (QFT) \cite{Calabrese:2009qy,Calabrese:2004eu}, it is hard to calculate the entanglement entropy of an interacting QFT. In this situation, the AdS/CFT correspondence asserts that a one-dimensional higher gravity theory corresponds to a strongly interacting QFT. Intriguingly, Ryu and Takayanagi (RT) showed how to calculate the entanglement entropy of a strongly interacting system on the dual gravity side \cite{Ryu:2006ef,Ryu:2006bv}. The background metric of the RT formula is static, so that there is no nontrivial time-dependence due to the time-translational symmetry. After that, Hubeny, Rangammani, and Takayanagi (HRT) further claimed that in order to calculate the holographic entanglement entropy in the time-dependent geometry, one has to exploit the covariant (or HRT) formulation rather than the RT formula due to breaking of the time translation symmetry \cite{Hubeny:2007xt}. The HRT formulation was used to investigate the entanglement entropy change in the thermalization process \cite{Liu:2013iza,Liu:2013qca} and in the eternal inflation of the dS boundary model \cite{Maldacena:2012xp,Fischler:2013fba,Koh:2020rti,Park:2020xho,Giataganas:2021cwg}. 

In the standard cosmology \cite{Kinney:2009vz,Baumann2018}, one can obtain various expanding universes relying on the involved matter.  Although an AdS space easily realizes the eternal inflation at the boundary, it is not easy to find the dual bulk geometry of universes expanding by a power-law. In order to realize the standard Friedmann-Lema\^{i}tre-Robertson-Walker (FLRW)  cosmology holographically, we need to take into account another holographic model called the braneworld model (or Randall-Sundrum model) \cite{Israel:1966rt,Randall:1999ee,Randall:1999vf,Chamblin:1999ya,Dvali:1998pa,Park:2000ga,Park:2020jio}. The braneworld model is described by two AdS bulk geometries bordering at a one-dimensional lower hypersurface which we call a braneworld. Depending on the cosmological constants of two bulk geometries and the tension of the braneworld, the braneworld moves in the radial direction perpendicular to the brane's worldvolume. In string theory, gauge fields and their supersymmetric partners live in the braneworld and then are identified with the open string's fluctuations. On the other hand, the graviton is described by a closed string that lives in the ten-dimensional bulk space. In the braneworld model, the brane has a delta-function potential which makes a zero-mode of graviton confined at the brane. As a consequence, all fundamental particles we detect in nature live in the braneworld. Due to this reason, we can identify the braneworld with the universe we live in. Interestingly, the braneworld model enables us to investigate the cosmological evolution of the universe holographically. It is worth noting that in the braneworld model the expansion of the universe is described by the junction equation, instead of the Einstein equation, which determines the radial motion of the braneworld in the bulk \cite{Randall:1999vf,Park:2000ga,Park:2020jio}.

In the standard cosmology, the expansion rate is determined by the matter contained in the universe \cite{Kinney:2009vz,Baumann2018}. When a four-dimensional universe contains uniformly distributed $(p-1)$-dimensional objects, the expansion rate is given by $a \sim \ta^{2/(p-2)}$ where $a$ and $\ta$ are a scale factor and cosmological time respectively. In this case, the dimension of the extended object is related to the equation of state parameter, $w =(1-p )/3$. To  describe the standard cosmology holographically, we need to understand how we can realize such matter in the braneworld model, In string theory, the fundamental matter in the brane is described by an open string attached to the brane \cite{Polchinski:1996na,Duff:1994an}. Therefore, the gravitational backreaction of open strings can describe massive particles in the braneworld. This geometry  was known as the string cloud geometry  \cite{Letelier:1979ej,Stachel:1980zr,Stachel:1980zs,Gibbons:2000hf,Herscovich:2010vr,Chakrabortty:2011sp,Chakrabortty:2016xcb}. In this work, we concern a more general geometry with a $p$-brane gas \cite{Park:1997dw,Park:1999gn,Park:1999xn}. Then, the string cloud geometry corresponds to a $p$-brane gas geometry with $p=1$. We show that the braneworld model in a $p$-brane gas geometry reproduces the standard cosmology caused by $(p-1)$-dimensional extended objects. In this $p$-brane gas geometry, we further study the time evolution of the entanglement entropy. We find that the entanglement entropy evolves by the volume law in the early time era and by the area law in the late time era. In this case, since the expansion rate of the universe relies on the involved matter, the time-dependence of the entanglement entropy also crucially depends on the contained matter.

Classical information carriers cannot be delivered faster than the light velocity, so we can define a cosmological or particle horizon in expanding universes \cite{Kinney:2009vz,Baumann2018}. The inside and outside of the cosmological horizon are called a visible and invisible universe, respectively. Since these two universes are causally disconnected, there is no correlation at the classical level. However, this is not the case if regarding quantum correlations. Since the quantum correlation is nonlocal, there is still nontrivial correlation between two causally disconnected universes. This leads to a nontrivial entanglement entropy across the cosmological horizon. We also investigate how the entanglement entropy between visible and invisible universes changes with time in expanding universes.

The rest of this paper is organized as follows. In Sec. 2, we begin with summarizing the standard cosmology and compare it with the braneworld cosmology defined in the $p$-brane gas geometry. In Sec. 3, we investigate the time-dependent entanglement entropy in the expanding universe with $(p-1)$-dimensional objects. In Sec. 4, we further look into the entanglement entropy between visible and invisible universes across the cosmological horizon. Lastly, we finish this work with some concluding remarks in Sec. 5.

\section{Holographic dual of the standard cosmology}

First, we briefly summarize the standard FLRW cosmology \cite{Kinney:2009vz,Baumann2018} for later comparison with the braneworld model. Assuming that an ideal gas satisfying $p = w \r$, where $\r$ and $p$ are its energy density and pressure, is uniformly distributed in the universe, the Friedmann equation 
\be 		\la{equation:ScaleFactor}
\ls \fr{\dot{a}}{a} \rs^2 = -  \fr{k}{a^2} + \fr{\k^2}{3} \fr{\r_0}{a^{3(1+w)}}  .
\ee
determines the scale factor $a(\ta)$ to be
\be
a (\ta) \sim \ta^{\fr{2}{3(1+w)}},
\ee
where $\ta$ is the cosmological time. 

In general, the scale factor crucially relies on the equation of state parameter. If the matter is a relativistic massless field, we call it radiation and its equation of state parameter is given by $w=1/3$. In the radiation-dominated era, the scale factor increases with time by $a \sim \ta^{1/2}$. If the universe is filled by non-relativistic massive particles instead of radiation, we call such a non-relativistic matter a dust with $w=0$. In the matter-dominated era, the scale factor increases by $a \sim \ta^{2/3}$. Lastly, if the universe has a positive constant vacuum energy without any matter, the universe expands exponentially by $a \sim e^{H \ta}$ which we call eternal inflation. During the eternal inflation, the vacuum energy has the equation of state parameter, $w=-1$. If we further take into account extended or solitonic objects like cosmic strings and domain walls, such extended objects allow the universe to expand with a different power. For example, the cosmic strings with $w = -1/3$ enforce the universe to expand linearly with time, $a \sim \ta$. On the other hand, the uniform distribution of domain walls, whose equation of state parameter is $w=-2/3$, makes the universe expand by $a \sim \ta^{2}$.

Now, we investigate the holographic dual of the standard cosmology \cite{Randall:1999ee,Randall:1999vf,Brax:2004xh,Bousso:2002ju,Gubser:1999vj,Kraus:1999it}. Assume that two $(d+1)$-dimensional bulk spaces are bordered by a braneworld, $d$-dimensional hypersurface. To obtain a smooth $(d+1)$-dimensional manifold, the metrics of two bulk spaces must be continuous at the braneworld. However, the derivatives of the metrics on both sides of the braneworld usually have different values. To avoid this mismatch, we have to introduce a delta function-like potential which is associated with the stress tensor of the braneworld \cite{Chamblin:1999ya,Balasubramanian1999,Park:2021nyc}. This prescription leads to the junction equation and determines the radial motion of the braneworld in the bulk. Although we are able to consider the braneworld model with two different bulk geometries, for convenience we focus on the same bulk geometries with a $Z_2$ symmetry. This implies that one bulk geometry is the mirror of the other.

Let us concern a $(d+1)$-dimensional AdS geometry with different boundary topologies. If we require  a time translation invariance in an AdS space, the topology of the $d$-dimensional AdS boundary is given by ${\bf R} \times \S_k$ where ${\bf R}$ and $\S_k$ indicate the temporal and $(d-1)$-dimensional spatial sections of the AdS boundary. If a matter field is uniformly distributed, the gravitational backreaction of the matter field deforms the AdS metric. The deformed metric, as will be seen in the next section, is given by
\be			\la{metric:WantedForm}
ds^2 = - \fr{r^2}{R^2}  f_k(r) \ dt^2 + r^2 d \S_k^2 + \fr{R^2}{ r^2 f_k(r) } dr^2 ,
\ee 
with the following blackening factor
\be			\la{ansatz:generalmetricfactor}
f_k (r) = 1 + k \fr{R^2}{r^2} - \fr{\cal B}{r^{\cal A}} .
\ee
For ${\cal A}<0$ the asymptotic geometry is not an AdS space anymore. From now on, we restrict the range of  ${\cal A}$ to be in ${\cal A}>0$ to obtain an asymptotic AdS geometry.

Since the braneworld can move in the radial direction, its radial position is generally given by a function of time, $r(t)$. Therefore, the cosmological time in the braneworld is related to the bulk time 
\be
d \ta^2 
= \lb   \fr{r^2 \, f_k  }{R^2}   -  \fr{R^2}{ r^2 \,  f_k}  \ls \fr{dr}{dt} \rs^2 \rb  dt^2 .
\ee 
Then, the induced metric in the braneworld becomes a FLRW-type metric
\be		 
ds_B = - d \ta^2 + r^2 \,  d \S_k^2 .   \la{Result:FLRWmetric}
\ee
Therefore, the radial position of the braneworld can be identified with the scale factor. Substituting the above metric ansatz \eq{ansatz:generalmetricfactor} into the junction equation, we finally obtain
\be		
\ls \fr{\dot{r}}{r} \rs^2  = \fr{\s^2 - \s_c^2}{4 (d-1)^2} - \fr{k}{r^2}   + \fr{{\cal B}}{ R^2 r^{{\cal A}}}  , \la{Result:Jeqgeneral1}
\ee
with a critical tension 
\be
\s_c = \fr{2 (d-1) }{R }  .
\ee
This junction equation determines the radial position of the braneworld. To an observer living in the braneworld, this junction equation is observed by the Friedmann equation. More precisely, assuming that the brane has the critical tension for $d=4$, the above junction equation reduces to 
\be		    	\la{result:FriedmannonBrane}
\ls \fr{\dot{r} }{r} \rs^2  = - \fr{k}{r^2}   + \fr{{\cal B}}{ R^2 r^{\cal A}}  .
\ee
If ${\cal A}$ is further related to the equation of state parameter 
\be         \la{Result:relationw}
{\cal A} = 3(1+w) ,
\ee
the junction equation is equivalent to the Friedmann equation \eq{equation:ScaleFactor} of the standard cosmology. In the next section, we will show how to derive\eq{Result:relationw} in the braneworld model.


Above, we showed that the braneworld model can realize the standard cosmology with the metric ansatz \eq{ansatz:generalmetricfactor} and identification \eq{Result:relationw}. We ask what the origin of the identification \eq{Result:relationw} is and how a bulk field is associated with the matter content of the standard cosmology. From now on, we discuss how we can obtain the identification \eq{Result:relationw} in the braneworld moving in the $p$-brane gas geometry. Before discussing a $p$-brane gas geometry, we first consider a black hole solution, which is one of the examples for a $p$-brane gas geometry. If a bulk field is localized at the center of the five-dimensional AdS space, it allows a black hole 
\be         \la{Ansatz:metricfactor}
f_k (r) = 1 + k \fr{R^2}{r^2} - \fr{m}{r^{4}} ,
\ee
where $m$ is the mass of the black hole. According to the holographic renormalization, the black hole mass is associated with the stress tensor of a boundary matter. The energy density and pressure are proportional to $N^2 m$ where $N$ is the rank of the gauge group. In this case, the $N^2$ dependence is associated with the degrees of freedom of an adjoint matter, like a gauge boson. In addition, the boundary stress tensor is traceless with $w=1/3$, so that the adjoint matter must be massless. These results indicated that the black hole mass is related to the energy of the the massless adjoint field living in the braneworld. The equation of state parameter $w=1/3$, as expected, corresponds to the value of the radiation in the standard cosmology \cite{Park:2020jio}.


Now, we move to a $p$-brane gas geometry. When $p$-branes are distributed in an AdS space, a $p$-brane gas geometry is governed by the following action
\be
S = \fr{1}{2 \k^2} \int d^{d+1} x \sqrt{- G} \ls {\cal R} - 2 \L \rs  + T_p  N_p  \int d^{p+1} \xi \sqrt{-h} \  \pa^{\a} x_{M}  \, h_{\a\b} \, \pa^{\b} x_{N} \, G^{MN}  ,
\ee
where $N_p$ and $T_p$ are the number and tension of $p$-branes. Here, $x^M$ ($M =0,\cdots,d$) and $\xi^{\a}$ ($\a =0,\cdots,p-1, d$) indicate coordinates of a bulk spacetime and brane's worldvolume, respectively. The variation of the action with respect to a bulk metric reduces to
\be
\d S &=& \fr{1}{2 \k^2} \int d^{d+1} x \sqrt{- G} \ls {\cal R}_{MN} - \half {\cal R} G_{MN} + \L G_{MN}  \rs  \d G^{MN}  \nn 
&& + T_p  N_p  \int d^{p+1} \xi \sqrt{-h} \  \pa^{\a} x_{M}  \, h_{\a\b} \, \pa^{\b} x_{N} \, \d G^{MN}  .
\ee
Note that the integral measure of the bulk has a different dimension from the one of the brane's worldvolume. Due to this reason, we cannot directly write the Einstein equation. To avoid this problem, we assume that $p$-branes are uniformly distributed in spatial directions perpendicular to the brane's worldvolume. Denoting the coordinates perpendicular to the worldvolume as $y^a$, $y^a$ becomes coordinates of a $(d-p)$-dimensional space with an appropriate metric $g_{ab}$. Now, we take a static gauge satisfying $\xi^{\a} = x^\a$ and $y^a= x^a$ with $\a=\lc 0, \cdots,p-1,d \rc$ and $a=\lc p, \cdots, d-1 \rs$ and denote the temporal and radial coordinates as $x^0=t$ and $x^d = r$. In this case, the background metric involving the gravitational backreaction of branes is given by 
\be			
ds^2 = \fr{r^2}{R^2}   \ls - f (r) \ dt^2 + \d_{ij} d x^i dx^j  \fr{}{}\rs + \fr{R^2}{r^2  f(r) } dr^2  ,
\ee 
where  $i=\lc 1, \cdots ,d-1 \rc$. To obtain an asymptotic AdS geometry, we require $f(\infty)=1$.

Above, the number of $p$-branes can be expressed in terms of the number density $\bar{n}_p$
\be
N_p = \int d^{d-p} y \sqrt{g} \ \bar{n}_p  ,
\ee 
where the integral measure indicates an integration over the perpendicular directions. It is worth noting that the number density $\bar{n}_p$ depends on the radial position because the perpendicular volume relies on the radial position. If we further rewrite $\bar{n}_p = n_p / \sqrt{g} = n_p \, R^{d-p}  / r^{d-p}$,  $n_p$ corresponds to a constant number density independent of the radial position. Under this parameterization, the integration over the brane worldvolume is rewritten as the integral over the bulk space
\be
\int d^{p+1}   \xi^a  \, \sqrt{- h } \, \int d^{d-p} {y} \, \sqrt{ g} = \int d^{d+1} x \sqrt{- G} .
\ee
Then, the variation of the $p$-brane's action is rewritten as
\be
\d S_p = \int d^{d+1} x \sqrt{-G} \  T_{MN} \, \d G_{MN}  ,
\ee
where the stress tensor of $p$-branes becomes 
\be
T_{MM}  = - \fr{T_p n_p \, R^{d-p} }{r^{d-p}} \lc G_{tt} ,  \fr{p-1}{d-1} G_{11} , \cdots,    \fr{p-1}{d-1}  G_{(d-1) \, (d-1)}  , G_{uu} \rc  ,
\ee
with $T_{MN}=0$ for $M \ne N$. Here, $(p-1)/(d-1)$ indicates the average number of $p$-branes extending to one of the spatial directions.

After resolving the issue on different integral measures, we can derive the Einstein equation
\be
R_{MN} - \half G_{MN} R + G_{MN} \L = 2 \k^2 T_{MN} .  
\ee
Substituting the metric ansatz into this Einstein equation and solving it finally determines the metric factor $f(r)$ to be
\be
f (r) &=& 1  -   \fr{\r_p}{r^{d-p}}  .
\ee
where $\r_p$ is related to the $p$-brane's energy density
\be
\r_p=  c_p  \, \k^2 n_p \, T_p \, R^{d-p+2}  .   \la{Relation:hodensity}
\ee
Here $c_p$ is an appropriate numerical number replying on $p$, for example, $c_1 = 4/3$, $c_2 = 2/3$, and $c_3 = 4/9$ for $d=4$. This was called the $p$-brane gas geometry \cite{Park:2020jio}. For $p=d$, a $d$-brane  fill up the bulk space so that the gravitational backreaction of $d$-branes modifies only the bulk cosmological constant. For $p=d-2$, intriguingly, the $p$-brane gas geometry results in an isotropic momentum relaxation geometry \cite{Park:2016slj}. More precisely, the scalar field in the momentum relaxation geometry is associated with the hodge dual of a $(p+1)$-form gauge field generated by the $p$-brane.

Assuming that the $p$-branes have the critical tension and comparing the junction equation derived in the $p$-brane gas geometry  
\be		
\ls \fr{\dot{r} }{r} \rs^2  = - \fr{k}{r^2}   + \fr{\r_p}{ R^2 r^{d-p}}  ,
\ee
with the Friedmann equation \eq{equation:ScaleFactor} for $d=4$, we see that the $p$-brane gas geometry leads to the following equation of state parameter
\be
w = \fr{1 - p}{3} .
\ee
This is equivalent to the identification assumed in \eq{Result:relationw}. As a consequence, the $p$-brane gas corresponds to $(p-1)$-dimensional objects in the braneworld. For example, the bulk $0$- and $1$-branes are dual of the radiation ($w=1/3$) and dust ($w=0$) in the braneworld. On the other hand, a $2$-brane gas gives rise to $w=-1/3$ which is the value of the cosmic string in the standard cosmology. For $p=3$, a $3$-brane gas reduces to a domain walls with $w=-2/3$. The resulting scale factor in the braneworld reduces to
\be				\la{Result:pscale}
a(\ta) = \ta^{2/(4-p)}  = \ta^{\fr{2}{3(w+1)}} ,
\ee
which is consistent with the scale factor of the previous standard cosmology. The braneworld shows accelerating expansion for $p>2$ and decelerating expansion for $p<2$.

\section{Entanglement entropy in the universe with extended objects}

Now, we look into how the entanglement entropy evolves in expanding universes. For $d=2$ with $p=0$ and $1$, the time-dependent entanglement entropy was investigated and compared with another holographic model called the dS boundary model \cite{Park:2020xho}. Here, we study the time-dependent entanglement entropy of the standard cosmology with matters. 

We consider the entanglement entropy contained in a disk-shaped region \cite{Nishioka:2009un,Nishioka:2018khk}. To parameterize the entangling surface which is the boundary of the entangling region, we introduce a new radial coordinate, $z=R^2/r$, and rewrite the $p$-brane gas geometry for $d=4$ as the following form
\be			
ds^2 = \fr{1}{z^2} \ls  - f(z) dt^2 +  du^2 + u^2 d s_{S^2}^2 + \fr{1}{ f(z) } dz^2  \rs,
\ee 
with
\be
f (z) = 1 - \r_p z^{4-p}  ,
\ee
where we set $R=1$ and $k=0$ for simplicity. Parameterizing the entangling region by
\be
0 \le u \le l  ,
\ee
the entanglement entropy governed by the minimal surface is given by
\be			\la{Action:HEE}
S_E = \fr{\O_2}{4 G} \int_0^l d u \fr{u^2 \sqrt{z'^2 + f(z)}}{z^3  \sqrt{f(z)}}  ,
\ee 
where $\O_2$ indicates a solid angle of a two-dimensional unit sphere. Notice that the subsystem size $l$ is the size measured by a comoving observer. The physical size in the expanding universe is given by $a(\ta) \, l$. Depending on the physical system we are interested in, we can take several different subsystem sizes. The first one is to take a constant $l$. In this case, the subsystem size measured by a comoving observer does not change but the physical size expands because the background space expands. The second case we will consider later is to identify the cosmological horizon with an entangling surface. The velocity of light is always finite, so that there exists the bound of a visible universe, the so-called  cosmological horizon. Since the cosmological horizon is time-dependent even to a comoving observer, we have to take a time-dependent subsystem size $l (\ta)$ to express the cosmological horizon appropriately.

We first take into account the case with a constant $l$. The entanglement entropy in the expanding universe with cosmic strings is described by the minimal surface extending to the $2$-brane gas geometry. For $p=2$, the equation of motion derived from \eq{Action:HEE} determines the configuration of the minimal surface 
\be
0 = z'' + \frac{2 \left(z'\right)^3}{u \left(1-\rho _2 z^2\right)}+\frac{\rho _2 z \left(z'\right)^2}{1-\rho _2 z^2} +\frac{3 \left(z'\right)^2}{z} +\frac{2 z'}{u} +\frac{3 \left(1-\rho _2 z^2\right)}{z} .
\ee
It is usually hard to find an analytic solution of this differential equation. Therefore, we take into account specific parameter regions allowing perturbation. To do so, we first introduce a turning point $z_t$. Then, the minimal surface extends to only the range of $\bar{z} \le z \le z_t < z_h (= 1/\sqrt{\r_2})$ where $\bar{z}$ and $z_h$ correspond to the radial position of the braneworld and horizon, respectively.  

We first focus on the case of $z_t \ll z_h$. When $\bar{z} $ is fixed, the turning point $z_t$ determines the subsystem size.  For $z_t \ll z_h$, the subsystem size becomes much smaller than the inverse temperature $1/T_H$. Therefore, this limit corresponds to a UV limit. In this UV limit, the configuration of the minimal surface can be evaluated by applying the following series expansion
\be
z (u) = z_0 (u) + \r_2 z_1 (u) + \cdots ,
\ee
where the ellipsis indicates higher order corrections. After substituting the expansion form into the equation of motion and solving it order by order, the leading solution is given by
\be
z_0 (u)  = \sqrt{ z_t^2- u^2}  .
\ee 
where $z_t = \sqrt{l^2 + \bar{z}^2}$.
When we obtained this solution, we impose two boundary conditions, $z_0 '(l)=0$ and $z_0 (l)=\bar{z}$. The first condition is required to obtain a smooth minimal surface at the turning point. On the other hand, the second condition implies that the minimal surface anchors to the entangling surface defined on the braneworld.

Now, we move to the first correction, $z_1 (u)$. Using the leading solution, the first correction is given by
\be
z_1 (u) &=& \frac{c_1 \left(z_t-u\right){}^{2}}{u \sqrt{z_t^2 - u^2}} + \frac{6 \left(c_2+z_t^4\right)+5  z_t^2 u^2 -u^4}{6 \sqrt{z_t^2-u^2}} \nn
    && -\frac{z_t^3 \left(2  z_t  u \log \left(z_t+u\right)+\left(z_t-u\right){}^2 \tanh^{-1}\left( \fr{u}{z_t} \right)\right)}{u \sqrt{\left(z_t-u\right) \left(z_t+u\right)}}  ,
\ee
where $c_1$ and $c_2$ are two integral constants. These integral constants can be fixed by imposing two natural boundary conditions. The first one is $z_1'(0)=0$ for the smooth minimal surface at $u=0$ which determines one of the integral constants to be
\be
c_1 =0.
\ee
The other boundary condition we must impose is $z_1 (l)=0$ because the entangling surface does not change even after the perturbation. This requirement fixes the remaining integral constant to be
\be
c_2 = \frac{1}{6} \left(l^4-5 l^2 z_t^2+12 z_t^4 \log \left(l+z_t\right)-6 z_t^4\right)+\frac{z_t^3 \left(z_t-l\right){}^2 \tanh
   ^{-1}\left(\frac{l}{z_t}\right)}{l} .
\ee
Using these integral constants, we finally obtain 
\be
z_1 (u) &=&  -\frac{z_t^3 \left(2 u z_t  \log \left(z_t+u\right)+\left(z_t-u\right){}^2 \tanh ^{-1}\left(\frac{u}{z_t}\right)\right)}{u \sqrt{z_t^2-u^2}} \nn
    && + \frac{l (l^2-u^2) \left(l^2-5 z_t^2+u^2\right)+6 z_t^3 \left(2 l z_t \log \left(l+z_t\right)+\left(z_t-l\right){}^2 \tanh
   ^{-1}\left(\frac{l}{z_t}\right)\right)}{6 l \sqrt{z_t^2-u^2 }} .
\ee
Plugging the obtained solutions into the entanglement entropy formula and performing the integral, the entanglement entropy results in
\be
S_E &=& \frac{\Omega _2}{8 G} \left[ \frac{  l \sqrt{l^2+\bar{z}^2}}{\bar{z}^2}-  \tanh^{-1}\left(\frac{l}{\sqrt{l^2+\bar{z}^2}}\right)    \right]   \nn
    && - \frac{\r_2  \Omega _2}{48 G}  \left[ \frac{8 l^3+6 l \bar{z}^2}{\sqrt{l^2+\bar{z}^2}}
    - 3 \left(l^2+\bar{z}^2\right) \lc 3  \log \left(\frac{\sqrt{l^2+\bar{z}^2}+l}{\sqrt{l^2+\bar{z}^2}-l}\right)
    - 4  \tanh^{-1}\left(\frac{l}{\sqrt{l^2+\bar{z}^2}}\right) \rc \right]   \nn
   && + {\cal O} \ls \r_2^2 \rs  .
\ee 

Above the subsystem size is determined as a function of $\bar{z}$ and $z_t$. For $\bar{z} \ll z_t \ll z_h$, the subsystem size is much bigger than $\bar{z}$ and the entanglement entropy reduces to
\be
S_E  \approx \frac{\Omega _2}{8 G} \fr{l^2 }{\bar{z}^2} +\frac{\Omega _2}{16 G} \ls 2 \log
   \fr{\bar{z}}{2 l}+1 \rs   - \frac{  \Omega _2  }{24 G} \ls 3 \log  \fr{\bar{z}}{2 l} + 4 \rs \r_2  l^2+ \cdots  .
\ee
In this case, the leading contribution comes from the short-range correlation satisfying the area law, $S_E \sim l^2 \Omega _2$. This UV feature universally appears regardless of the dimension of the $p$-brane. For $p=3$, for example, the leading entanglement entropy again shows the similar area law
\be
S_E &\approx& \frac{ \Omega _2}{8 G } \fr{l^2}{\bar{z}^2} +\frac{\Omega _2}{16 G} \left( 2  \log \fr{\bar{z}}{2 l} + 1 \right)   + \frac{ \Omega _2}{8 G} \fr{l^2 \rho _3}{ \bar{z}}+  \cdots .
\ee
Despite this universal feature, the different expansion rate relying on the $p$-brane gas leads to a different time-dependence. Recalling that the brane position in braneworld model is inversely proportional to the scale factor, $\bar{z} = 1/a(\ta) = 1/ \ta^{2/(4-p)}$, the small value of $\bar{z}$ corresponds to the large scale factor in the late time era. In the expanding universe, as a result, the entanglement entropy in the late time era increases by
\be               \la{Result:lateHEE}
S_E \sim \frac{l^2 \Omega _2}{8 G}  \ta^{4/(4-p)}  .
\ee

For $\bar{z} \to z_h$ with satisfying $\bar{z} < z_t < z_h$, the previous perturbation is not valid. Therefore, we have to exploit another method to consider this parameter range. According to Ref. \cite{Kim:2016jwu}, when $z_t$ the entanglement entropy \eq{Action:HEE} for $z_t \to z_h$ is rewritten as
\be		 
S_E = \fr{\O_2}{4 G z_t^3} \int_0^l du u^{2} + \fr{\O_2}{4 G z_t^3}  \int_0^l d u \fr{u^2 \ls  z_t^3 \sqrt{z'^2 + f(z)}   -  z^3 \sqrt{f(z)}   \rs}{z^3  \sqrt{f(z)}}  .
\ee
Near the turning point ($u = 0$ and $z=z_t$ with $z' =0$), since the second term is negligible, the first term leads to the main contribution. Recalling that the volume of the subsystem is given by 
\be
V = \O_2 \int_0^l du u^{2} = \fr{\O_2 l^3}{3},
\ee 
the leading contribution becomes approximately
\be
S_E \approx S_{BH}&=& \fr{1}{4 G} \fr{V }{\bar{z}^{3} } + \cdots ,
\ee
where $\bar{z} \approx z_t$ and the ellipsis means small quantum corrections. Due to $\bar{z} = 1/a(\ta) = 1/ \ta^{2/(4-p)}$ in the braneworld model, the large $\bar{z}$ indicates a small scale factor in the early time era. Therefore, the entanglement entropy in the early time era satisfies the volume law  
\be			\la{Result:earlyHEE}
S_E \approx \frac{ \Omega _2 l^3}{12 G}  \ta^{6/(4-p)}  .
\ee

In summary, the braneworld model shows how rapidly the entanglement entropy grows in expanding universes. When $(p-1)$-dimensional objects are uniformly distributed, the universe expands with the scale factor $a(\ta) =\ta^{2/(4-p)}$ as shown in \eq{Result:pscale}. In this expanding universe, the entanglement entropy increases by the volume law in the early time era ($S_E \sim \ta^{6/(4-p)}$), whereas it in the late time era grows by the area law ($S_E \sim \ta^{4/(4-p)}$).

\section{Entanglement entropy across the cosmological horizon}

When the universe expands, we can see only the inside of the cosmological horizon because the outside is causally disconnected. Due to this reason, we call the inside and outside of the cosmological horizon a visible and invisible universe, respectively. In expanding universes, the cosmological horizon usually grows up with time. Although we cannot get any classical information from the invisible universe, it is not the case for quantum theory. Since quantum entanglement is nonlocal, there still exist nontrivial quantum correlation between two causally disconnected regions. Therefore, it would be interesting to study the quantum entanglement between the visible and invisible universes by identifying the cosmological horizon with the entangling surface. 

In the expanding universe described by the FLRW metric, the distance traveled by light during the time interval, $\D \ta = \ta-\ta_i$, is defined as \cite{Kinney:2009vz,Baumann2018}
\be			\la{def:cosmodis} 
d_c (\ta) = a (\ta) \int_{\ta_i}^{\ta} \fr{ d \ta'}{a(\ta')} ,
\ee
where $\ta_i$ is an appropriate initial time and the light speed sets to be $c=1$. From now on, we assume for simplicity that the cosmological horizon at the initial time ($\ta_i=0$) is located at zero ($d_c (\ta_i)=0$). Then, $d_c (\ta)$ corresponds to the cosmological horizon measured at present time $\ta$. For an expanding universe by a power-law, the cosmological horizon results in  
\be
d_c(\ta) \sim \fr{4-p }{2-p} \ta  .
\ee
This indicates that the cosmological horizon in the power-law expansion increases linearly with time, regardless of the expansion power. In addition, the positivity of $d_c (\ta)$ restricts the value of $p$ to be in the range of $p < 2$. If $p > 2$, there is no cosmological horizon. More precisely, the scale factor and cosmological horizon for $p=2$ behave like $a (\ta) \sim \ta$ and $d_c (\ta) \sim \ta  \log \fr{\ta}{\ta_i}$. On the other hand, the scale factor for $p >2$ behaves like $a (\ta) \sim \ta^a$ with $a>1$ and the cosmological horizon is given by
\be
d_c (\ta) = \ta^a \ls \ta^{1-a} - \ta_i^{1-a} \rs < 0  .
\ee
Since the cosmological horizon must be positive, it is not well defined for $p > 2$. This is because the expansion rate of universe is faster than the light velocity.

We now identify the cosmological horizon with an entangling surface to calculate the entanglement entropy between the visible and invisible universes. Since an observer living at the center of the visible universe cannot get any classical information from the invisible universe, it is natual to identify the cosmological horizon with an entangling surface. Recalling that we defined the subsystem size $l$ in the comoving frame, we need to know where the cosmological horizon appears in the comoving frame. Using \eq{def:cosmodis}, the cosmological horizon in the comoving frame appears at
\be
l (\ta) = \fr{d_c (\ta)}{ a(\ta)}   .
\ee
For a power-law expansion, the subsystem size must be time-dependent to describe the cosmological horizon correctly
\be		\la{Result:sizepowerexp}
l (\ta)  \sim   \fr{4-p}{2-p} \ \ta^{1-2/(4-p)} .
\ee
When we identify the cosmological horizon with the entangling surface, the area of the minimal surface is again given by \eq{Result:earlyHEE} and \eq{Result:lateHEE} with the time-dependent subsystme size \eq{Result:sizepowerexp} rather than the constant one used in the previous section. Therefore, the leading entanglement entropy across the cosmological horizon increases in the early time era 
\be
S_E \sim  \ta^{3} .
\ee
In the late time era, the entanglement entropy grows by 
\be
S_E \sim  \ta^{2} .
\ee
These results show that, although the expansion rate crucially relies on the matter content, the entanglement entropy between visible and invisible universes, regardless of the matter content for $p<2$, increases by $\ta^3$ in the early time era and by $\ta^2$ in the late time era. For $p=2$, the entanglement entropy across the cosmological horizon increases by $\ls \ta \log \ta \rs^3$ in the early time and by $\ls \ta \log \ta \rs^2$ in the late time era.

\section{Discussion}

We have studied the time-dependent entanglement entropy of expanding universes in the braneworld model. The braneworld cosmology, unlike the standard cosmology, is determined by the junction equation rather than the Einstein equation. More precisely, the junction equation determines the brane's radial motion, which is detected by the expansion of the universe to an observer living in the braneworld. We took into account $p$-branes uniformly distributed in the AdS space and found the $p$-brane gas geometry involving the gravitational backreaction of $p$-branes. When $p$-branes extend to the radial direction, the warping factor of the background AdS space allows a black hole-type geometry. Since an observer living in the braneworld cannot see the bulk's radial direction, he or she detects $p$-branes as $(p-1)$-dimensional extended objects. When $(p-1)$-dimensional objects are uniformly distributed in the standard cosmology, the scaling behavior of the spatial coordinate determines the equation of state parameter, for example, $w=0$ for the dust ($p=1$), $w=-1/3$ for the cosmic string ($p=2$), and $w=-2/3$ for the domain wall ($p=3$). Intriguingly, we showed that the braneworld model in the $p$-brane gas geometry reproduces the exactly same equation of state parameters. This indicates that the braneworld model can holographically realize the standard cosmology.

In general, it is hard to calculate the entanglement entropy in the expanding universe even for a free QFT, because the time-dependent background geometry makes the equation of motion complicated. In this work, however, we showed how to calculate the time-dependent entanglement entropy of various expanding universes in the braneworld model. When the subsystem size in the comoving frame is fixed, we showed that the leading entanglement entropy in the early time era evolves by
\be
S_E \sim \ta^{6/(4-p)} ,
\ee 
whereas it in the late time era increases by
\be
S_E \sim \ta^{4/(4-p)} .
\ee
This features indicate that the entanglement entropy increases by the volume law in the early time era and by the area law in the late time era.

In the expanding universe, the visible universe which an observer can see is restricted due to the finiteness of the light velocity. In this case, the cosmological horizon naturally appears as a border of the visible and invisible universes. Since these two universes are causally disconnected at the classical level, there is no classical correlation. However, if we further concern quantum correlations, the entanglement entropy across the cosmological horizon does not vanish due to the nonlocality of the quantum correlation. Since an observer in the visible universe cannot get any information from the invisible universe, we can identify the cosmological horizon with an entangling surface. We studied how the entanglement entropy across the cosmological horizon evolves in the expanding universes. Intriguingly, the time evolution of the entanglement entropy across the horizon shows a different behavior from that of the expanding universe with a fixed comoving distance. We showed that the cosmological horizon  in the four-dimensional spacetime is not well defined for $p>2$. For $p < 2$, we found that the entanglement entropy, regardless of the matter content, increases in the early time era by
\be
S_E \sim \ta^3 ,
\ee
while it in the late time era increases by
\be
S_E \sim \ta^2 .
\ee
For $p=2$ which corresponds to cosmic strings in the braneworld, however, we found that the entanglement entropy grows up in the early time era by
\be
S_E \sim \ls \ta \log \ta \rs^3 ,
\ee
and in the late time era  by
\be
S_E \sim \ls \ta \log \ta \rs^2 .
\ee
It would be interesting to investigate how such quantum entanglement entropy affects the cosmological history and structure formation. We hope to report more interesting results on this issue in future works. \\

{\bf \large Acknowledgement} 

This work was supported by the National Research Foundation of Korea(NRF) grant funded by the Korea government(MSIT) (No. NRF-2019R1A2C1006639).




%

\end{document}